\documentclass[twocolumn]{aastex6}
\usepackage{amsmath}
\newcommand{\NA}{---}
\pdfoutput=1
\makeatletter
\newcommand*{\rom}[1]{\expandafter\@slowromancap\romannumeral #1@}
\makeatother

\slugcomment{}

\shorttitle{Improved Constraints on the Disk Around MWC 349A from the 23-Meter LBTI}
\shortauthors{Sallum et al.}

\begin{document}

\title{Improved Constraints on the Disk Around MWC 349A from the 23-Meter LBTI}

\author{S. Sallum\altaffilmark{1}, J. A. Eisner\altaffilmark{1}, P. M. Hinz\altaffilmark{1}, P. D. Sheehan\altaffilmark{1}, A. J. Skemer\altaffilmark{2}, P. G. Tuthill\altaffilmark{3}, J. S. Young\altaffilmark{4}}
\altaffiltext{1}{Astronomy Department, University of Arizona, 933
  N.\ Cherry Ave., Tucson, AZ 85721, USA}
\altaffiltext{2}{Astronomy Department, University of California Santa Cruz, 1156
  High St., Santa Cruz, CA 95064, USA}
\altaffiltext{3}{School of Physics, University of Sydney, Sydney, NSW 2006, Australia}
\altaffiltext{4}{Cavendish Laboratory, University of Cambridge, J J Thompson Avenue, Cambridge, UK}

\email{email: ssallum@email.arizona.edu}

\begin{abstract}
We present new spatially resolved observations of MWC 349A from the Large Binocular Telescope Interferometer (LBTI), a 23-meter baseline interferometer made up of two, co-mounted 8-meter telescopes. 
MWC 349A is a B[e] star with an unknown evolutionary state.
Proposed scenarios range from a young stellar object (YSO), to a B[e] supergiant, to a tight binary system.
Radio continuum and recombination line observations of this source revealed a sub-arcsecond bipolar outflow surrounding a $\sim100$ mas circumstellar disk.
Followup infrared studies detected the disk, and suggested that it may have skew and an inner clearing.
Our new infrared interferometric observations, which have more than twice the resolution of previously-published datasets, support the presence of both skew and a compact infrared excess.
They rule out inner clearings with radii greater than $\sim 14$ mas.
We show the improvements in disk parameter constraints provided by LBTI, and discuss the inferred disk parameters in the context of the posited evolutionary states for MWC 349A.

\end{abstract}

\section{Introduction}\label{sec-intro}

Discovered in 1932 as a member of a binary system \citep{1932ApJ....76..156M}, MWC 349A is a B[e] star with an uncertain spectral type  \citep[e.g.][]{1998A&A...340..117L,Allen:1972b}.
It lacks optical photospheric lines; however, He \rom{1} emission indicates a high stellar temperature \citep{1996A&AS..118..495A}.
Estimates range between $20,000 - 35,000$ K, corresponding to B0 \citep{2002A&A...395..891H} to late O \citep{1980ApJ...239..905H} spectral type.
Its mass and luminosity determinations range from 30 $-$ 40 M$_{\odot}$  \citep[e.g.][]{Ponomarev:1994,Planesas:1992,2012A&A...541A...7G,2013A&A...553A..45B} and $3\times10^4$ $-$ $8 \times 10^5$ L$_{\odot}$ \citep{1985ApJ...292..249C,2012A&A...541A...7G}, respectively.
Its distance may be as close as 1.2 kpc, based on the spectral type for MWC 349B \citep{1985ApJ...292..249C}, or as large as 1.7 kpc \citep{2002AJ....123.1639M,2000A&A...360..539K} if A is not associated with B \citep[e.g.][]{2013ApJ...777...89S,2012A&A...541A...7G,2002AJ....123.1639M} and is instead a member of the Cyg OB2 association.

MWC 349A is one of the brightest radio sources in the sky \citep{1972Natur.240..230B} and exhibits masing emission from the far-infrared through the millimeter \citep{1989A&A...215L..13M,Thum:1994,Strelnitski:1996b,Thum:1998}.
Continuum observations at 6.1 cm reveal a sub-arcsecond nebula with a dark lane roughly 100 mas wide at its equator \citep[e.g.][]{1985ApJ...292..249C,White:1985,1993ApJ...418L..79M}.
The radio spectrum indicates an ionized wind expanding at $25-50$ km s$^{-1}$ \citep{1981A&A....93...48A}, yielding an inferred mass loss rate of $10^{-5}$ M$_{\odot}$ per year \citep{1975A&A....39..217O}.

Spectroscopic and spectropolarimetric observations suggest the presence of a disk with both an ionized and a neutral component around MWC 349A \citep{1980ApJ...239..905H,1996A&A...312..234Y,1986ApJ...311..909H,1990MNRAS.247..466A,1977ApJ...218..170T}.
The maser emission supports this; double peaked line profiles indicate Keplerian rotation of gas \citep{1992A&A...256..507T,1992ApJ...387..701G,Ponomarev:1994}.
H92$\alpha$ line observations reveal rotation in the bipolar outflow and constrain its inclination to be $15\pm5^\circ$ with respect to the plane of the sky \citep{1994ApJ...428..324R}.
Assuming the disk and outflow are perpendicular, this suggests that the disk may be nearly edge-on.
The H30$\alpha$ recombination line originates from two locations consistent with the size and orientation of the nebula's dark lane \citep{Planesas:1992}, suggesting that the disk may reside there.

The disk characteristics inferred from radio data agree with high-resolution infrared imaging.
Early speckle observations constrain the disk size to be smaller than the dark lane in the radio \citep{1983A&A...120..237M}.
Gaussian fits to subsequent speckle imaging yield best fit FWHMs of $38\pm18$ mas in the north-south direction at K band, and $85\pm19$ mas in the east-west direction at L$'$ \citep{1986A&A...155L...6L}.
More recent interferometric observations can be modeled by uniform ellipses with similar sizes at wavelengths from 1.65 to 3.08 $\mu$m \citep{2001ApJ...562..440D}.
The reconstructed 1.65 $\mu$m image appears asymmetric \citep{2001ApJ...562..440D}.
Emission from the inner rim of an inclined disk with a clearing \citep[e.g.][]{2001Natur.409.1012T} or forward scattered light from a significantly flared disk \citep[e.g.][]{1998A&A...337..832K} could have caused this asymmetry.

MWC 349A has an unknown evolutionary state.
The presence of a dusty disk, infrared excess \citep{1970ApJ...161L.105G, Allen:1972b,Allen:1972,1973MNRAS.161..145A}, and bipolar outflow indicate a YSO morphology \citep{1977ApJ...218..170T,1985ApJ...292..249C}.
Recent observations associate it with a nearby cold molecular cloud, supporting this scenario \citep{2013ApJ...777...89S}.
While its binarity is uncertain, MWC 349B is a B0 III star, and an evolved companion would argue against a YSO morphology.
Proposed alternate scenarios to a YSO include a B[e] supergiant \citep[e.g.][]{1980ApJ...239..905H, 2002A&A...395..891H}, a binary system with an equatorial stellar wind \citep{1981ApJ...249..572M}, and a runaway hierarchical triple \citep{2012A&A...541A...7G}.

Here we present new infrared interferometric observations of the MWC 349A disk from the 23-meter Large Binocular Telescope Interfrerometer (LBTI).
We fit geometric and radiative transfer models to, and reconstruct images from the observations.
We compare the constraints on disk parameters derived from both the single-aperture (up to 8 meter baselines within each LBT primary mirror) and dual-aperture (baselines between the two primaries up to 23 meters) datasets.
We demonstrate the degeneracies in reconstructing images from sparsely sampled observations and emphasize the importance of applying both model fitting and imaging to these datasets.
We discuss the implications of the observations for the disk morphology and the evolutionary state of MWC 349A.

\section{Technique}\label{sec-tech}
Non-redundant masking \citep[e.g.][]{2000PASP..112..555T} transforms a filled aperture into an interferometer via a pupil plane mask.
The detector records the interference fringes formed by the mask, which we Fourier transform to calculate complex visibilities. 
From the complex visibilities we calculate squared visibilities, the powers on all baselines, and closure phases, sums of phases around baselines forming a triangle \citep[e.g.][]{1958MNRAS.118..276J,1986Natur.320..595B}.
Closure phases are intrinsically self-calibrating and are robust to atmospheric phase noise.
Since closure phases are correlated we project them into linearly independent combinations of closure phases \citep[e.g.][]{2013MNRAS.433.1718I,2015ApJ...801...85S} called kernel phases \citep{2010ApJ...724..464M}.
Due to the loss of phase information intrinsic to the technique we use model fitting and image reconstruction to understand the source brightness distribution.

Although NRM blocks the majority of incident light, it provides a much better point spread function characterization than a conventional telescope.
This enables imaging at smaller angular separation than more traditional direct imaging techniques such as filled-aperture angular differential imaging \citep[e.g.][]{2006ApJ...641..556M} and coronography \citep[e.g.][]{2014ApJ...780..171G}. 
While coronagraphs create inner working angles of $\sim \lambda/D$ for the highest performance designs \citep[e.g.][]{2005ApJ...633.1191M}, NRM provides resolution even within the diffraction limit.
It has proven useful in the direct detection of close-in stellar \citep[e.g.][]{2012ApJ...753L..38B,2008ApJ...678L..59I} and substellar \citep[e.g.][]{2015Natur.527..342S,2012ApJ...745....5K} mass companions. 

\section{Observations}
\begin{figure}
\epsscale{1.0}
\plotone{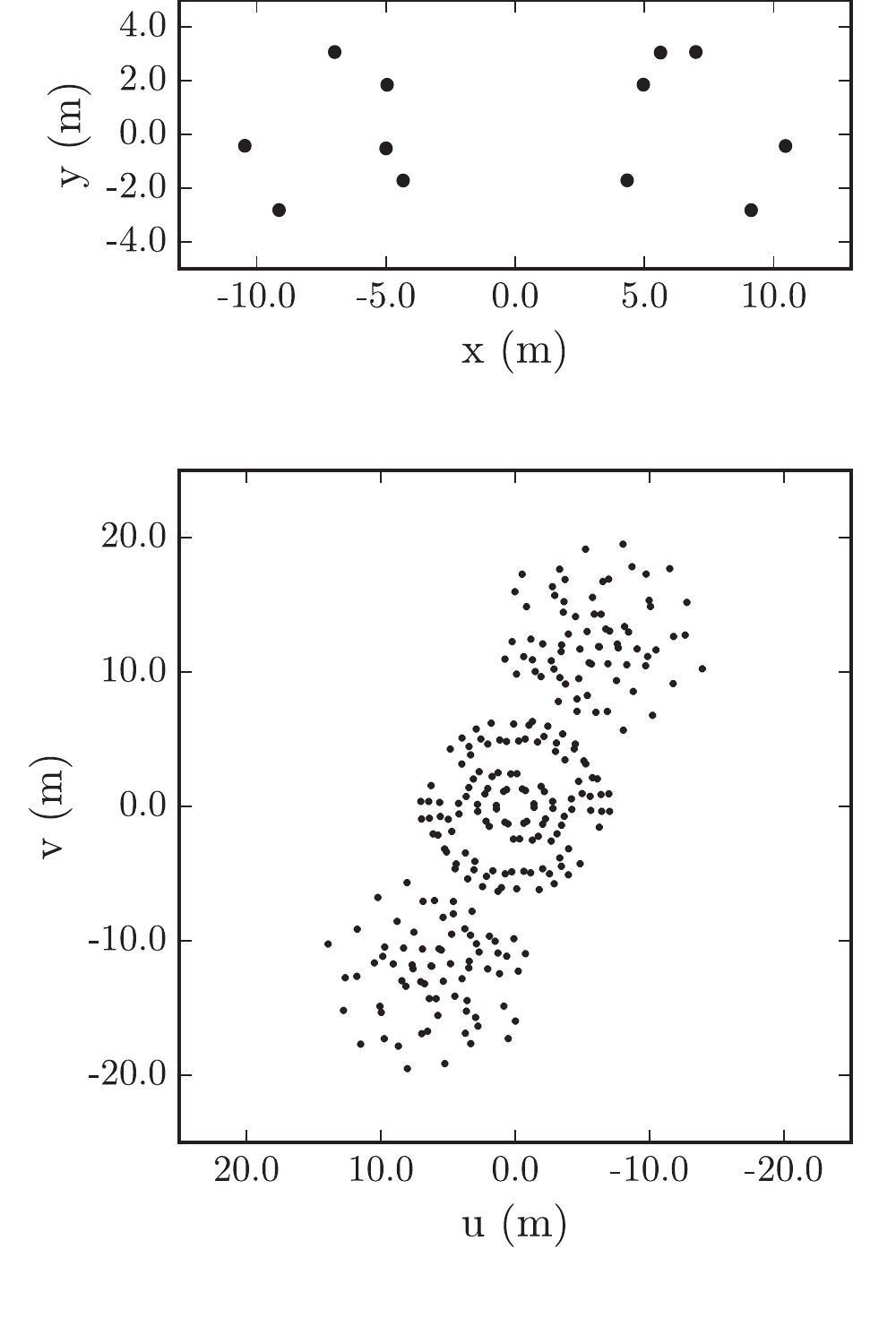}
\caption{\textbf{Top:} 12-hole mask installed in LBTI/LMIRCam. \textbf{Bottom:} Fourier coverage of the MWC 349 observations. The small amount of sky rotation means that some position angles were sampled with higher resolution than others.}
\label{fig-mask}
\end{figure}

We observed MWC 349A on 21 May 2012 at the LBT with the 12-hole mask (see Figure \ref{fig-mask}) installed in LBTI/LMIRCam \citep{2008SPIE.7013E..28H,2012SPIE.8446E..4FL}.
This configuration provided baselines up to $\sim 23$ meters and yielded 66 squared visibilities and 220 closure phases that we projected into 55 independent kernel phases. 
We took data with the adaptive optics correction running on each of the two LBT apertures.
We did not actively correct the path length between them to enable long exposures for the baselines connecting the two mirrors.
We rather aligned them once at the beginning of the night and took short enough exposures for the long baselines to be coherent.

To account for instrumental signals, we observed the unresolved calibrator star HD 193092 with the same configuration as MWC 349A. 
We used a bandpass centered on 3.78 $\mu$m with a width of 0.2 $\mu$m.
The dataset for each object consists of two cubes of 500 29-ms exposures, yielding 29 seconds of total integration.
Each cube of images was taken with even sampling over a time interval of 145 seconds with a 0.27 second dead time between frames.  
The two MWC 349A datacubes were taken at LST (HA) of 19h 18m (-1h 12m) and 19h 40m (-0h 52m), resulting in $\sim 13^\circ$ of sky rotation (see Figure \ref{fig-mask}). 

\section{Data Reduction}

We flat field, sky subtract, and bad pixel correct all images, then Fourier transform them to form complex visibilities. 
The non-zero mask hole size and bandpass cause information from each baseline to be encoded in several pixels  in the Fourier transform (``splodges"). 
To calculate squared visibilities, we sum the power in the splodges corresponding to each baseline and normalize by the power at zero baseline.
We subtract the average power in the regions without signal to correct for any bias, then average the squared visibilities for all individual images to calculate the squared visibility for each cube of images.
To calculate closure phases, for each triangle of baselines we find all pixel combinations that satisfy the following relation:
\begingroup\makeatletter\def\f@size{8.5}\check@mathfonts
\begin{equation}
(u_1,v_1)+(u_2,v_2)+(u_3,v_3) = 0.
\label{eq-close}
\end{equation}
\begingroup\makeatletter\def\f@size{10.0}\check@mathfonts
and multiply their complex visibilities to form a bispectrum.
We calculate the bispectra for all pixel triangles that connect the three splodges and satisfy Equation \ref{eq-close}.
We average these to form the bispectrum for each triangle of baselines for a single image.
We then average the bispectra for all images and take the bispectral phase as the closure phase for each triangle of baselines.
We lastly project the closure phases into kernel phases.

Since we have only two calibrator observations, we simply average the mean kernel phases and squared visibilities for the two data cubes. 
We subtract the calibrator kernel phases from the target kernel phases, and divide the target visibilities by the calibrator visibilities. 
Since calibration errors introduce the largest amount of scatter in the final kernel phases and visibilities, we would normally use the scatter in a large number of calibrator scans to estimate the errors for the target observations \citep[e.g.][]{2015Natur.527..342S}.
However,  we cannot robustly estimate errors using only two calibrator measurements.
Thus we assume that the errors are uniform and take the kernel (closure) phase errors to be the standard deviation of all calibrated kernel (closure) phases.
We similarly take the standard deviation of all squared visibilities after subtracting the two dithers from each other to remove any trends.\footnote{The calibrated closure phases and squared visibilities can be found at \url{www.stephsallum.space/research/MWC349A}.} 
This results in a kernel (closure) phase error of $3.4^\circ~(6.0^\circ)$, and a squared visibility error of 0.08. 
These values agree with those derived when we include uniform error scalings as nuisance parameters in the fitting (\S $\ref{sec-fitting}$).  

\section{Model Fitting and Image Reconstruction}\label{sec-fitting}

\subsection{Geometric Models}

To estimate the size of the MWC 349A disk, we first fit uniform ellipses to the calibrated kernel phases and squared visibilities.
This model is identical to that published in \cite{2001ApJ...562..440D}: a solid ellipse with semi-major axis $R_{out}$, position angle $\theta$ measured east of north, and axial ratio $r$.
Depending on the disk inclination and geometry, a bright inner disk rim, gas or refractory dust within the sublimation radius, or the central star may be visible.
We thus also fit geometric models that include central delta functions accounting for a fraction $b$ of the total flux, beginning with a uniform ellipse plus delta function model.
These two models are symmetric and cannot cause non-zero kernel phase measurements.
Since the asymmetry in the 1.65 $\mu$m Keck image could have resulted from forward scattering from a flared disk, we also consider skewed ellipse models.
The skewed ellipse is the uniform ellipse multiplied by a sinusoid in position angle, given by the following \cite[e.g.][]{2010AJ....140.1838S}:
\begingroup\makeatletter\def\f@size{8.5}\check@mathfonts
\begin{equation}
   \label{eq-ell}
    I=
    \begin{cases}
      1+A_s \cos\left(\phi_s - \phi\right), & \text{if}\ \left(\frac{x^{\prime}}{R_{out}}\right)^2 + \left(\frac{y^{\prime}}{r R_{out}}\right)^2 < 1 \\
      0, & \text{otherwise}
    \end{cases}
\end{equation}
\begingroup\makeatletter\def\f@size{10.0}\check@mathfonts\\
where
\begingroup\makeatletter\def\f@size{8.5}\check@mathfonts
\begin{equation}
\begin{split}
x^{\prime} = x \cos\left(\theta\right) - y \sin\left(\theta\right)\\
y^{\prime} = x \sin\left(\theta\right) + y \cos\left(\theta\right)\\
\phi = \arctan\frac{y}{x}.
\end{split}
\end{equation}
\begingroup\makeatletter\def\f@size{10.0}\check@mathfonts\\
\noindent Here $\phi_s$ is the position angle at which the flux is brightest.
Given the high temperature and luminosity estimates for MWC 349A, a clearing in the dust disk may be resolved.
We thus also fit skewed ring plus delta function models to allow for a compact component (the star plus any gaseous / refractory material within the sublimation radius) and an outer disk. 
The skewed ring model is the skewed ellipse with an inner hole of radius $R_{in}$: 
\begingroup\makeatletter\def\f@size{8.5}\check@mathfonts
\begin{equation}
   \label{eq-ring}
    I=
    \begin{cases}
      1+A_s \cos\left(\phi_s - \phi\right), & \text{if}\ \left(\frac{x^{\prime}}{R_{out}}\right)^2 + \left(\frac{y^{\prime}}{r R_{out}}\right)^2 < 1 \\
       0, &\text{if}\ \left(\frac{x^{\prime}}{R_{in}}\right)^2 + \left(\frac{y^{\prime}}{r R_{in}}\right)^2 < 1\\
      0, & \text{otherwise}
    \end{cases}
\end{equation}
\begingroup\makeatletter\def\f@size{10.0}\check@mathfonts
where $x^\prime, y^\prime, \phi, \text{and}~\phi_s$ are identical to those in Equation \ref{eq-ell}. 

We also fit two dimensional Gaussians to the data to explore models without sharp edges.
Like the solid ellipse fits, we first consider simple Gaussians and then add a central delta function and skew.
The skewed Gaussian brightness profile is given by the following:
\begingroup\makeatletter\def\f@size{8.5}\check@mathfonts
\begin{equation}
\begin{aligned}
    I = & ~\left(1+A_s \cos\left(\phi_s - \phi\right)\right) \\ 
         &\times \exp \left[ -4 \ln{2} \left( \left( \frac{x'}{rHWHM} \right)^2 + \left(\frac{y'}{HWHM} \right)^2\right) \right]
\end{aligned}
\end{equation}
\begingroup\makeatletter\def\f@size{10.0}\check@mathfonts\\
where $x'$, $y'$, and $\phi_s$ are defined in the same way as Equation \ref{eq-ell}.
We lastly fit Gaussian ellipses with inner clearings to the data.
In order to make the simpler Gaussians a subset of these models, to make the ring model we start with a simple non-skewed Gaussian.
We then subtract a second Gaussian with identical position angle and axis ratio, but with $HWHM$ scaled by $f_{HWHM}$. 
We constrain $f_{HWHM}$ to be less than 1 to prevent negative signal in the model images.
We lastly apply skew and add a central delta function.

We fit the data using the Markov chain Monte Carlo algorithm \texttt{emcee} \citep{2013PASP..125..306F}. 
We apply parallel tempering to ensure that the parameter space is fully explored in the case of multiple likelihood maxima.
We calculate the 1$\sigma$ parameter errors using the 16\% and 84\% contours from the chains at a temperature of one.
To compare the various models, we calculate the Bayesian evidence \citep[e.g.][]{2008ConPh..49...71T}, the integral of the posterior probability over the parameter priors, or the probability of a model given the data. 
The evidence ratios, or log evidence differences, between two models give their relative probabilities.
Since Bayesian evidence is a noisy statistic with a non-zero false positive probability \citep[e.g.][]{2011MNRAS.413.2895J}, we also compute $\chi^2$ differences to compare the models.
For each model we calculate the difference between its minimum $\chi^2$ value and that of the most complex model with fewer parameters.

We fit the data once including the kernel phase and visibility error scalings as nuisance parameters.
Since the best fits were nearly identical to the measured scatter we present results where we fix the error scalings to the observed kernel phase and visibility scatter.
We also perform fits to the intra-aperture baselines to understand how the full LBT resolution improves the model parameter constraints.

\begin{deluxetable*}{lccccccc}
\tabletypesize{\scriptsize}
\tablecaption{Geometric Model Fit Results \label{tbl-fits}}
\tablewidth{0pt}
\startdata
\\
\multicolumn{8}{c}{Ellipse Models}\\
\hline
Model &$R_{out}$ (mas)& $\theta$ ($^\circ$) &r &A$_s$ &$\phi_s$ (deg) &$b$ & $R_{in}$ (mas)   \\
\hline
Ellipse					&$46\pm{2}$				&$97 \pm{3}$ 			&$0.65 \pm0.03$					&\NA													&\NA							&\NA								&\NA								\\
Ellipse + $\delta$	&$57 \pm{2}$				&$99 \pm{3}$ 			&$0.66 \pm0.03$					&\NA													&\NA							&$0.30\pm{0.02}$					&\NA						\\
Ellipse + $\delta$ + Skew &$58 \pm2$					&$98 \pm{3}$		 	&$0.68 \pm0.03$					&$0.17\pm0.04$								&$-153\pm^7_6$		&$0.32\pm0.01$					&\NA					\\
Ring + $\delta$	 + Skew &$57\pm2$					&$98 \pm3$				&$0.68\pm0.03$					&$0.16\pm0.04	$								&$-153\pm^7_6$		&$0.33\pm{0.02}$					&$< 14$		\\
\hline
\multicolumn{8}{c}{Gaussian Models} \\
\hline
Model  &$HWHM$ (mas)&$\theta$ ($^\circ$) &r & A$_s$ &$\phi_s$ (deg) & $b$   & $f_{HWHM}$\\
\hline
Gaussian								&$28.2 \pm0.7$				&$97\pm3$ 				&$0.64\pm0.03$					&\NA								&\NA										&\NA										&\NA\\
Gaussian + $\delta$				&$ 34 \pm 1$					&$101\pm3$ 			&$0.64\pm0.03$					&\NA								&\NA										&$0.23\pm^{0.02}_{0.03}$	&\NA\\
Gaussian + $\delta$	 + Skew&$ 34 \pm 1$					&$101\pm3$			 	&$0.66\pm0.03$					&$0.24\pm0.02$					&$-153\pm^7_5$								&$0.24\pm0.02$					&\NA\\
Gaussian Ring + $\delta$ + Skew		&$32\pm^{2}_{3}$				&$101\pm^{4}_{3}$				&$0.67\pm^{0.03}_{0.04}$			&$0.21\pm0.06$ 		&$-153\pm^6_8$	&$ 0.4 \pm0.2$	 &	$0.31\pm^{0.03}_{0.04}$ 	
\enddata
\end{deluxetable*}
\renewcommand{\arraystretch}{1.0}

\begin{deluxetable*}{lcccccc}
\tabletypesize{\scriptsize}
\tablecaption{Model Comparison\label{tbl-mods}}
\tablewidth{0pt}
\tablehead{\colhead{Model} & \colhead{$\chi^2_{min}$} & \colhead {d.o.f.} & \colhead{$\Delta\chi^2$}\tablenotemark{a} & \colhead{$\Delta$d.o.f.}\tablenotemark{a} &  Significance\tablenotemark{b} & \colhead{$\log{Z}$}}
\startdata
Ellipse & 264.5 & 239 &\NA & \NA &\NA &$-139\pm2$\\
Ellipse + $\delta$ & 234.4& 238 & 30.1& 1 & $5.5\sigma$ &$-126\pm3$\\
Ellipse + $\delta$ + Skew & 213.9 & 236 & 20.5 &2& $4.1\sigma$ &$-120\pm3$\\
Ring + $\delta$ + Skew & 213.9 & 235 & 0.0 & 1 & \NA &$-121\pm4$\\
\hline
Gaussian &255.5 & 239 &\NA &\NA&\NA& $-134\pm2$\\
Gaussian  + $\delta$& 228.3 & 238 & 27.2 &1 &$5.2\sigma$ & $-123\pm3$\\
Gaussian + $\delta$ + Skew & 208.8 & 236 &19.5 & 2 & $4.0 \sigma$ & $-117\pm3$\\
Gaussian Ring + $\delta$ + Skew & 208.5 & 235 & 0.3 & 1& $<1\sigma$ & $-118\pm3$ \\
\enddata
\tablenotetext{a}{With respect to the above, simpler model}
\tablenotetext{b}{Derived from the $\chi^2$ difference test}
\end{deluxetable*}

\begin{figure*}
\epsscale{1.2}
\plotone{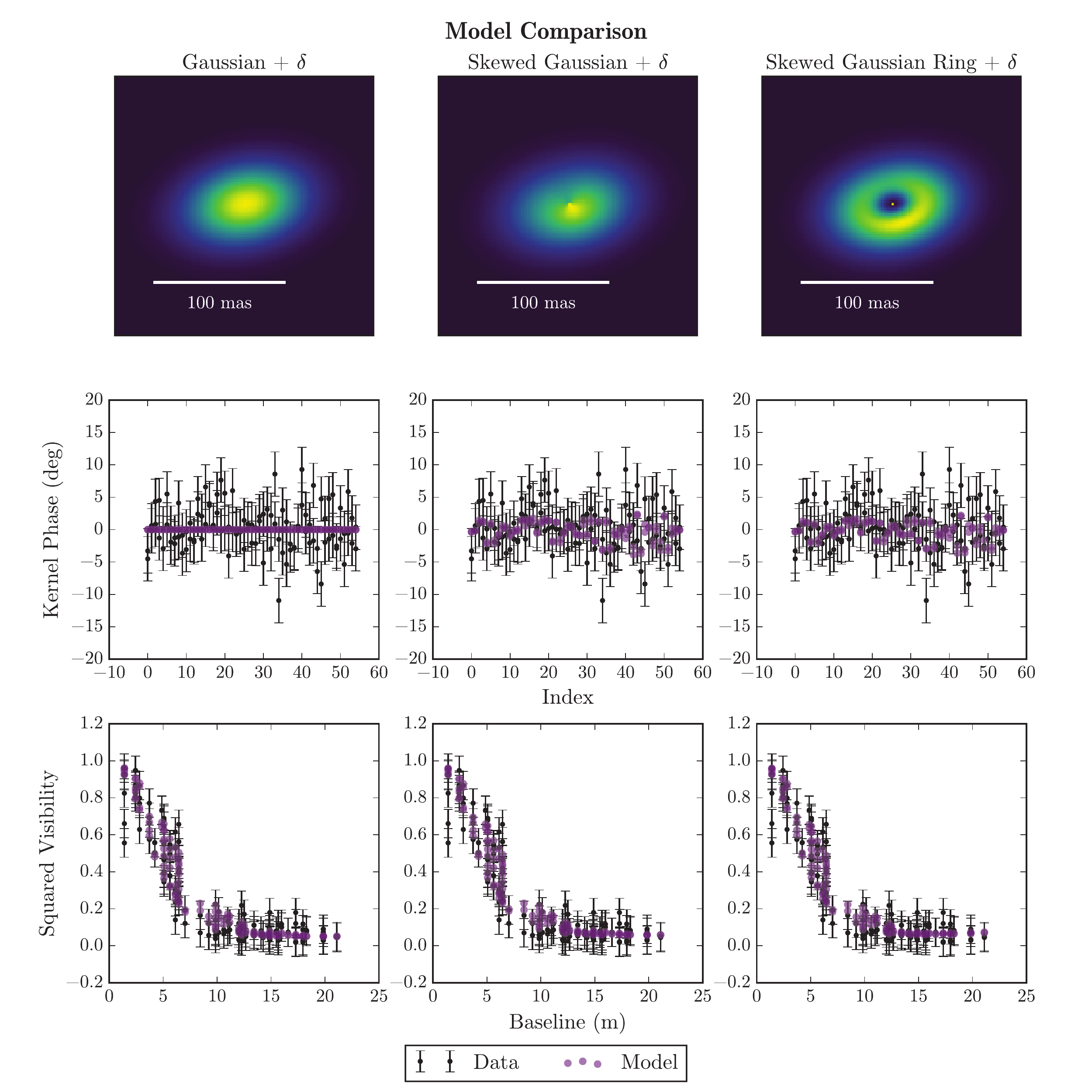}
\caption{Gaussian + $\delta$ (left column), Gaussian + $\delta$ + Skew (center column), and Gaussian Ring + $\delta$ + Skew (right column) model comparison. The black points show the observed kernel phases (middle row) and squared visibilities (bottom row), while the purple points show the model observables. These correspond to the last three models listed in Table \ref{tbl-fits}.}
\label{fig-mod_comp}
\end{figure*}

Table \ref{tbl-fits}  lists the best-fit dual-aperture model parameters and Table \ref{tbl-mods} lists their corresponding minimum $\chi^2$ and Bayesian evidence values.
The best fit position angles agree for all models and are also consistent with the best fit position angle reported in  \citet{2001ApJ...562..440D}.
For both types of brightness distributions, the Bayesian evidence and $\chi^2$ difference testing suggest that models including a compact component and skew are significantly better than the simpler models (see Table \ref{tbl-mods}).
These models provide a better match to the observations (see Figure \ref{fig-mod_comp}).

Both the Bayesian evidence and the  $\chi^2$ difference testing suggest that including an inner clearing does not improve the fit significantly.
The Ring + $\delta$ + Skew model constrains any inner hole to have a radius less than 14 mas, but the best fit is indistinguishable from the Ellipse + $\delta$ + Skew model, given the resolution of the observations.
While the Gaussian Ring + $\delta$ + Skew model has a slightly lower minimum $\chi^2$ than Gaussian models without an inner clearing, its $\Delta\chi^2$ is low enough that it is not preferred at the $1 \sigma$ level.
It produces nearly identical observables to the Gaussian + $\delta$ + Skew best fit model (see Figure \ref{fig-mod_comp}).
Its evidence value is also comparable to the Gaussian + $\delta$ + Skew best fit model.

Figure \ref{fig-tri} shows the posterior distributions for the Ring + $\delta$ + Skew model fit using both the intra- and dual-aperture observations.
The 23-meter LBTI places new and tighter constraints on all of the disk parameters compared to the single-aperture observations.
The uniform ellipse model fit to the dual-aperture data results in comparable parameter errors as previous Keck studies \citep{2001ApJ...562..440D}, but with $\sim 21\%$ the number of squared visibilities and $\sim 6\%$ the number of closure phases.

\begin{figure*}
\epsscale{1.2}
\plotone{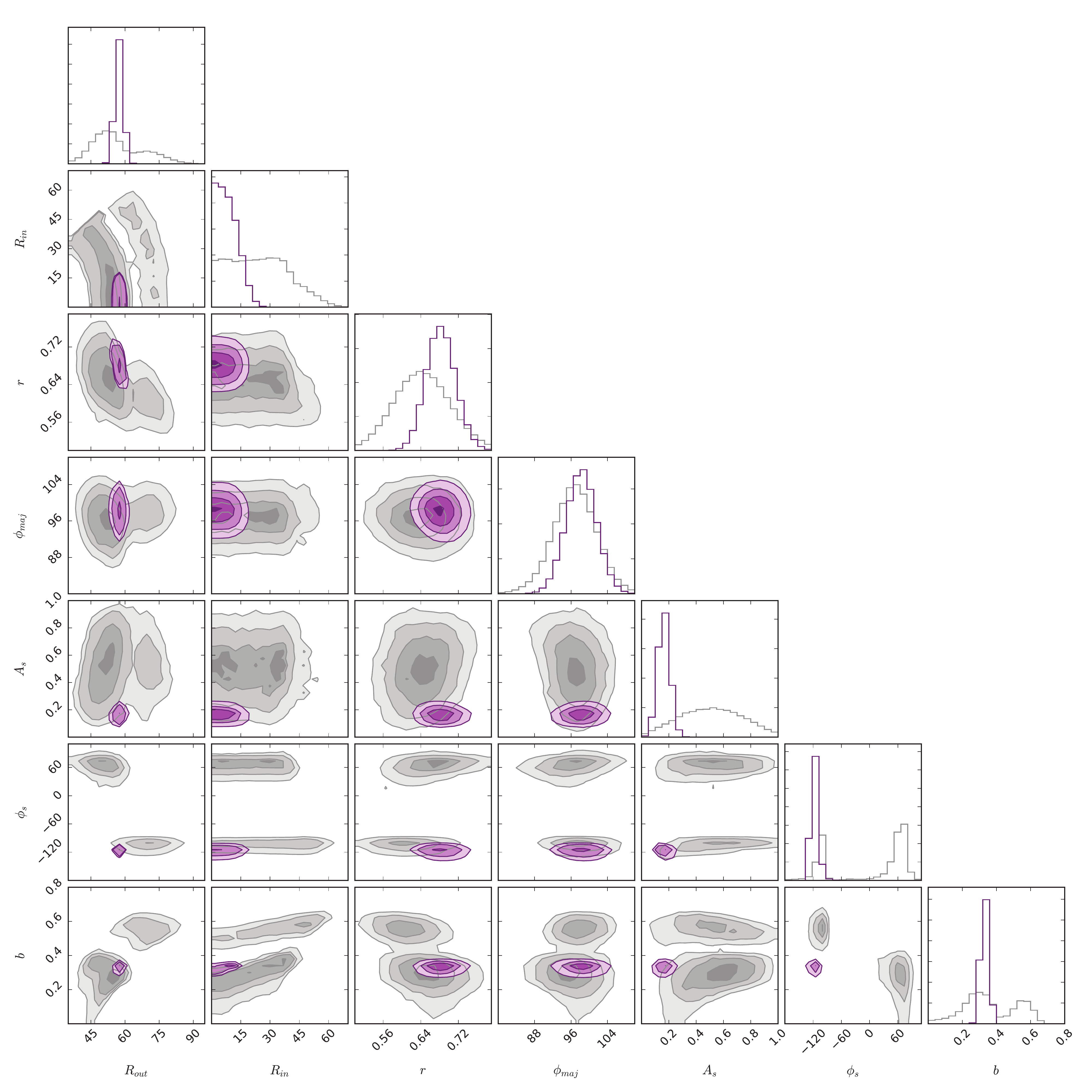}
\caption{Triangle plot for Ring + $\delta$ + Skew model fits. The grey and purple histograms at the top of each column show one-dimensional posterior distributions from the intra- and dual-aperture fits, respectively. The contours show joint posterior distributions for each pair of parameters in the model.}
\label{fig-tri}
\end{figure*}

\subsection{Radiative Transfer Modeling}
We generate radiative transfer models to test whether a disk in radiative equilibrium with the central star can match the observations.
We use the open source radiative transfer codes \texttt{Hyperion} \citep{2011A&A...536A..79R} and \texttt{RADMC-3D} \citep{2012ascl.soft02015D} and input the standard density profile for a flared disk:
\begin{equation}
\rho\left(r,z\right) = \rho_0\left(\frac{r}{r_0}\right)^{-\alpha} \exp\left(-\frac{1}{2}\left[\frac{z}{h\left(r\right)}\right]^2\right),
\end{equation}
where
\begin{equation}
h\left(r\right) = h_0 \left(\frac{r}{r_0}\right)^\beta.
\end{equation}
Here $r$ and $z$ are the radius and height in a cylindrical coordinate system.
The radius value $r_0$ is where the scale height $h$ is fixed to the constant value $h_0$ and the midplane density $\rho$ is fixed to the constant value $\rho_0$.
The density constant, $\rho_0$, can be found by integrating the density over all space with knowledge of the total disk mass.
We first consider scale height ($\beta$) and density ($\alpha$) power law indices (1.25 and 2.25, respectively) consistent with irradiated disks in hydrostatic equilibrium \citep[e.g.][]{2003ApJ...591.1049W,1998ApJ...500..411D}.
We set the disk inner radius at the point where the dust temperature reaches 1500 K to simulate dust sublimation, and use silicate dust with a grain size of $\sim1~\mu$m \citep[e.g.][]{2012ApJ...758..100B}.
We show results with a disk mass of 0.01 $M_*$, but also explored $10^{-3}~M_*$ and $0.1~M_*$ disk masses and found that they produce comparable results. 
We vary the stellar temperature and luminosity within their estimated uncertainties ($20,000 - 35,000$ K for temperature and $3\times10^4 - 8\times10^5$ L$\odot$ for luminosity.)
We set the scale height to outer disk radius ratio at 0.01, and the disk inclination to $75^\circ$ \citep{1994ApJ...428..324R}. 
We also explore models with higher flaring indeces, since MWC 349A may have a centrifugally driven disk wind \citep[e.g.][]{2011A&A...530L..15M}.

None of the radiative transfer models for passive irradiated disks match the observations. 
Figure \ref{fig-radmcres} shows two example disk models for the upper and lower bounds on the temperature and luminosity for MWC 349A.
For a low-luminosity MWC 349A, reprocessed light from the inner disk rim can account for the unresolved component in the geometric models. However, in this case the outer regions of the disk are too cold to produce significant amounts of emission. 
A high-luminosity MWC 349A is bright on the correct scales along the disk major axis, but due to its inclination it cannot reproduce the visibilities along the minor axis.
Asymmetric emission from the vertical wall at the disk inner edge also leads to a large phase signal.

\begin{figure}
\epsscale{1.2}
\plotone{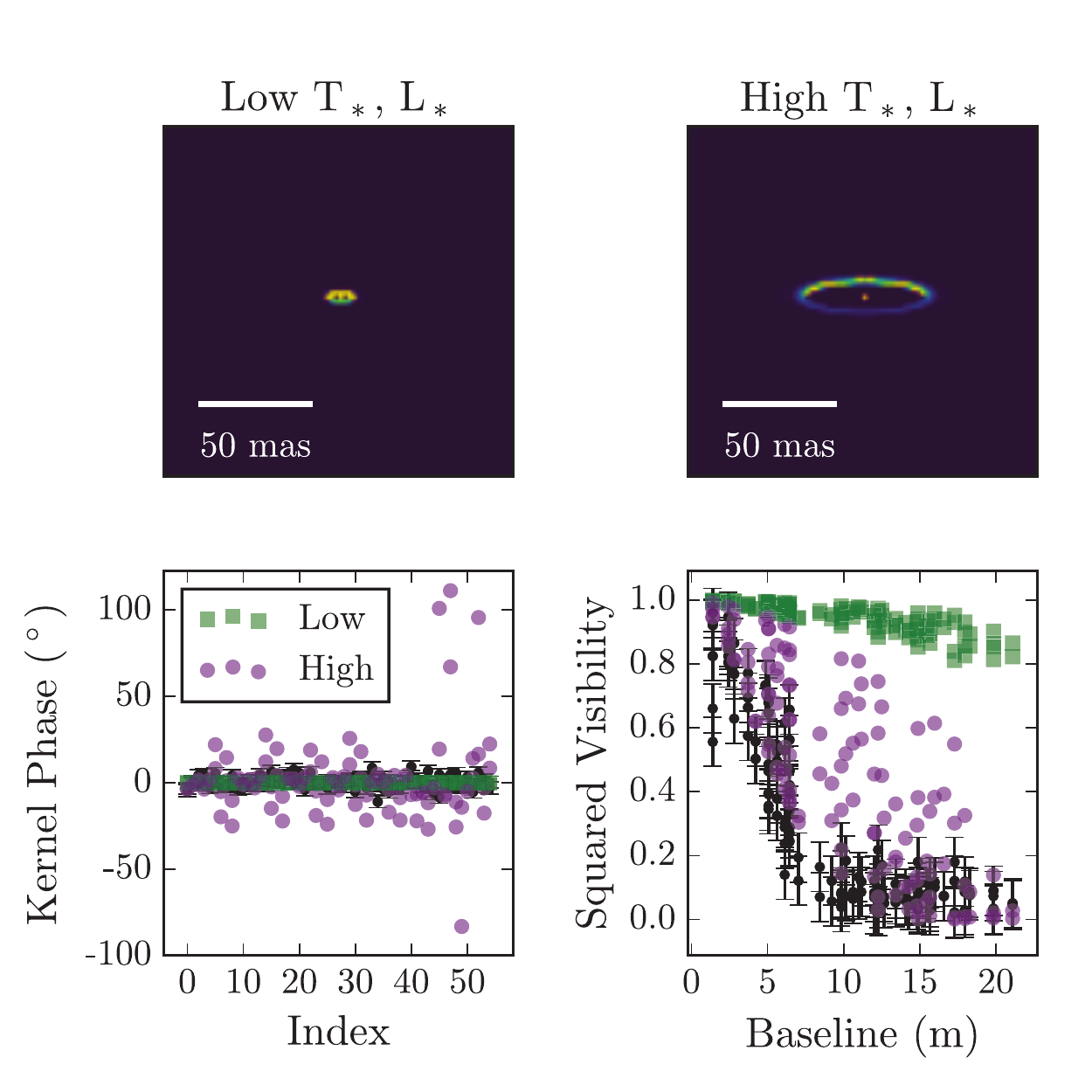}
\caption{\textbf{Top:} Ray-traced images for passive irradiated disk models using the lower (left) and upper (right) limits for MWC 349A's stellar temperature and luminosity. Both images have been rotated so the disk major axis is aligned with the x axis. \textbf{Bottom:} Kernel phases (left) and squared visibilities (right) for the model images, with the lower stellar luminosity model in green and the higher stellar luminosity model in purple. Black points with error bars show the observations.}
\label{fig-radmcres}
\end{figure}

\subsection{Image Reconstruction}

We reconstruct images using the \texttt{BSMEM} algorithm \citep{1994IAUS..158...91B}, assigning uniform closure phase and squared visibility errors of 6.0$^\circ$ and 0.08, respectively.
Degeneracies exist between different reconstructed images from datasets with sparse $(u,v)$ coverage and small amounts of sky rotation.
To illustrate this, we reconstruct images from simulated observations of the best fit model images.
We use the same $(u,v)$ coverage and sky rotation and add Gaussian noise at the level measured in the data.
We then reconstruct images from both the data and the simulations using multiple priors.

\begin{figure}
\epsscale{1.1}
\plotone{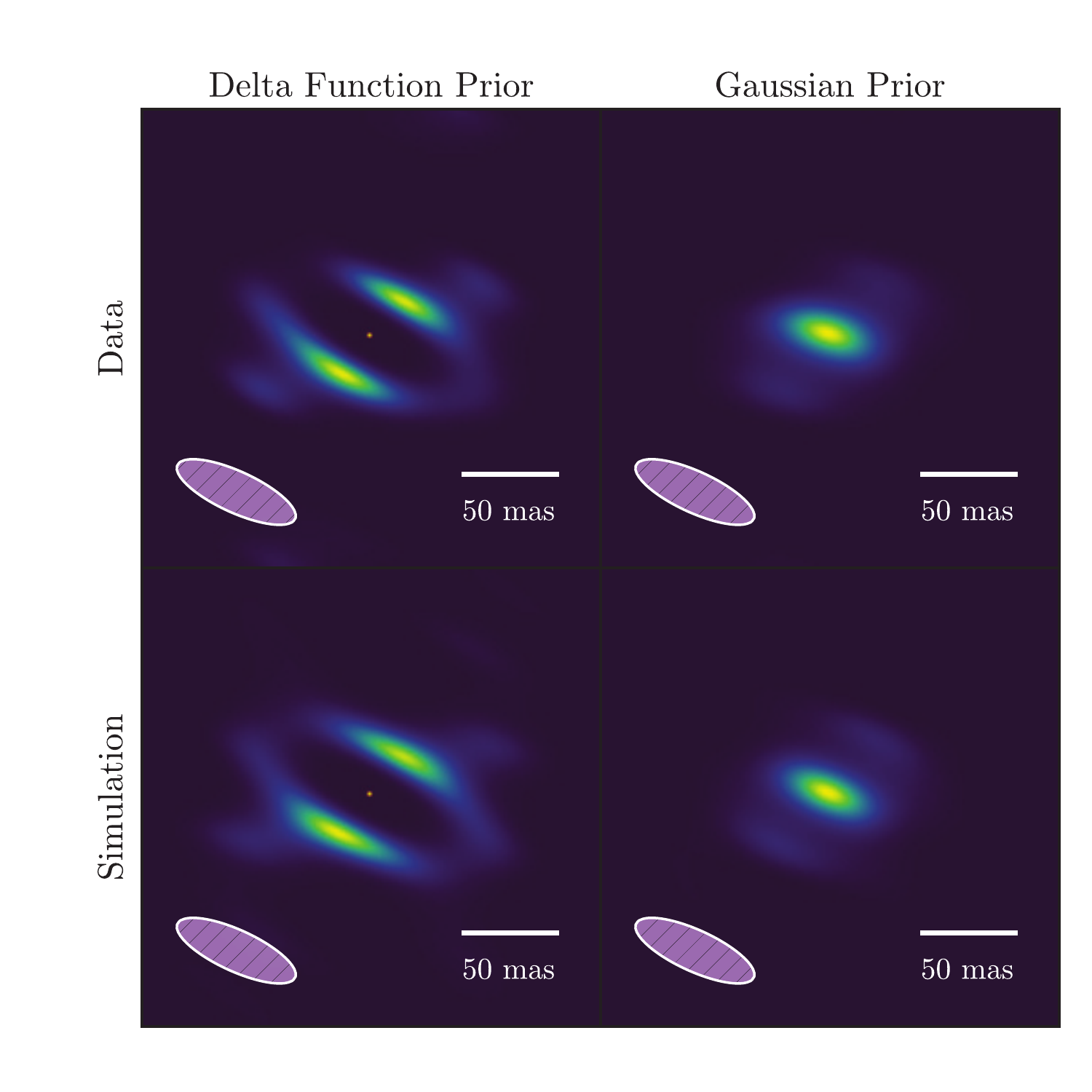}
\caption{Reconstructed images for MWC 349A observations (top row) and simulated observations of the best-fit skewed ring plus delta function model shown in Figure \ref{fig-mod_comp} (bottom). The left column shows images reconstructed using a delta function prior, and the right a Gaussian prior. The half-maximum contour of the synthesized beam is shown in the bottom left corner of each panel. The inability to reproduce the input image and the dependence on prior image highlight the need for model fitting and make reconstructed images ambiguous.}
\label{fig-recons}
\end{figure}

Figure \ref{fig-recons} shows images reconstructed from both the data and simulated observations of the best fit skewed ring plus delta function model.
Comparing the two rows of Figure \ref{fig-recons} shows that the best fit geometric model is consistent with images reconstructed using both priors. 
Comparing the two columns of Figure \ref{fig-recons} shows that the reconstructed images depend on the choice of prior image.
Additionally, degeneracies exist in the unresolved regions of the reconstructed images. 
The size and shape of the bright central component in each ``Gaussian Prior" image is consistent with the size and shape of the synthesized beam. 
The fractional flux contained in the central component is roughly the same for the images made using each prior, and is approximately equal to the amount of flux contained within the synthesized main beam in the input model image.
Putting a fraction $b$ of the image flux into a central component will create identical closure phases and squared visibilities as long as the central component is unresolved.
These degeneracies and the dependence on the prior image make reconstructed images ambiguous and necessitate model fitting in order to understand the source brightness distribution.

\section{Discussion}

\begin{figure}
\epsscale{1.2}
\plotone{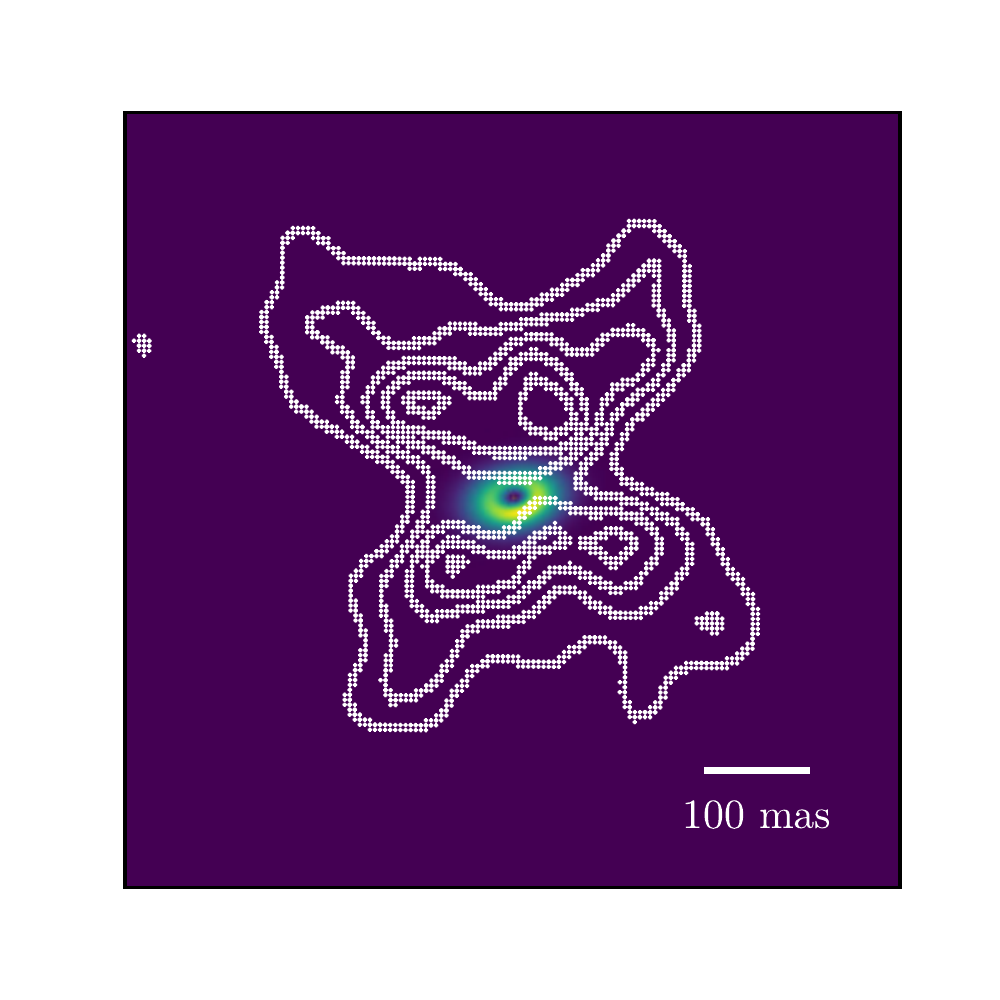}
\caption{Best fit Gaussian Ring + $\delta$ + Skew model shown with the VLA continuum map contours \citep{1993ApJ...418L..79M}. 
The position angle of the disk agrees with the orientation of the dark lane in the VLA map.}
\label{fig-composite}
\end{figure}

\subsection{Compact Infrared Excess}\label{sec-compact}
The compact component in the geometric models accounts for $\lesssim30\%$ the total image flux. 
Assuming a 3.78 $\mu$m flux of $\sim 100$ Jy for MWC 349A \citep{1977ApJ...218..170T} yields a 30 Jy flux for the central component.
Following \citet{2001ApJ...546..358M}, we can use the observed MWC 349A V and L band fluxes to estimate the amount of compact infrared excess. 
The emission expected for a star at temperature T with radius $\text{R}_\star$ is the Planck function times the solid angle, $\Omega = \frac{\pi \text{R}_\star^2}{d^2}$, where d is the distance to the star.
Using a dereddened V flux of 37.7 Jy and attributing it entirely to the star implies stellar radii of 13 - 28 $R_{\odot}$ depending on the chosen temperature and distance values.
With this range of stellar solid angles and temperatures, the amount of unextincted stellar flux expected at 3.78 $\mu$m is then $1 - 3$ Jy.
Thus at least $\sim90\%$ of the compact flux is in excess, and this estimate increases if we include extinction when calculating the stellar flux at 3.78 $\mu$m. 

Emission from a disk rim can account for the compact infrared excess if the stellar luminosity is low ($3\times10^4$ L$_\odot$) and the disk rim is close enough to the star to be unresolved. 
A higher luminosity ($8\times10^5$ L$_\odot$) MWC 349A sets the inner disk radius at $\sim40$ AU in the absence of shielding (Figure \ref{fig-radmcres}). 
This is highly resolved and cannot contribute to a compact infrared excess.
If the luminosity is indeed as high as $8\times10^5$ L$_\odot$, material such as optically thick gas \citep[e.g.][]{2009ApJ...692..309E} or refractory dust \citep[e.g.][]{2010A&A...511A..74B} must exist within the theoretical dust sublimation radius to explain the compact infrared excess.
Thus the central geometric model component could be caused by a close-in inner disk rim, material within the dust sublimation radius, or some combination of the two.

The inferred stellar radius and compact infrared excess are consistent with both the YSO and B[e] supergiant scenarios for MWC 349A.
Comparable stellar radii have been inferred for Herbig Ae/Be stars and B[e] supergiants \citep[e.g.][]{2006ASPC..355..135Z,2015MNRAS.453..976F}.
Observations of B[e] supergiants suggest compact infrared excesses with comparable fractional flux to that for MWC 349A \citep[e.g.][]{1986A&A...163..119Z,2012A&A...537A.103K}.
Large infrared excesses are found in observations of Herbig Ae/Be stars, in which the excess fractional flux can reach $95\%$ \citep[e.g.][]{2001ApJ...546..358M}. 
Symmetric gaseous emission has been detected within the dust sublimation radius of several Herbig Ae/Be stars \citep[e.g.][]{2009ApJ...692..309E,2010ApJ...718..774E,2008ApJ...677L..51T,2008ApJ...676..490K}.
This emission is $\lesssim$AU sized and consistent with the size of the compact component in the geometric models, given the distance estimates for MWC 349A. 
Gaseous emission coming from within the sublimation radius, which may be required if the stellar luminosity is high, would thus support an early age for MWC 349A.

\subsection{Disk Geometry}
The range of outer radii for the geometric disk models is $44-60$ mas, corresponding to $53-102$ AU given the MWC 349A distance uncertainties.
This is smaller than the gravitational radius for a photoevaporating disk, at which material would no longer be bound and could be lost in an outflow \citep{1994ApJ...428..654H}.
The gravitational radius can be written $r_g = GM_* / c_s^2$, where G is the gravitational constant, $M_*$ the stellar mass, and $c_s$ the sound speed.
Assuming $c_s = 11$ km s$^{-1}$ \citep{2001ApJ...562..440D}, $r_g = 219-290$ AU depending on the assumed stellar mass.
The best fit outer radii and position angles also agree with radio observations of the bipolar outflow and maser emission.
The H30$\alpha$ maser emission spots are separated by 65 mas \citep{Planesas:1992} at a position angle of $107\pm7^\circ$.
The best fit model is consistent with the width and orientation of the dark lane seen in VLA data as well  \citep[see Figure \ref{fig-composite};][]{1993ApJ...418L..79M}.
Thus the geometric model fits are consistent with a disk bound to MWC 349A at the center of the bipolar nebula and with the same orientation as the two maser spots.

The best fit ellipse size in \citet{2001ApJ...562..440D} increases with wavelength from a major axis of 36 mas at 1.65 $\mu$m to 62 mas at 3.08 $\mu$m.
These ellipse sizes, as well as the best fit major axis presented here ($88-120$ mas at $3.78~\mu$m) follow a wavelength scaling close to $\lambda^\frac{4}{3}$.
This trend is expected for flat, geometrically thin accretion disks as opposed to the $\lambda^2$ relation expected for flared disks \citep[e.g.][]{1995A&AS..113..369M}.
Without complete radiative transfer models to compare to the data, \citet{2001ApJ...562..440D} interpreted this as evidence for a flat disk around MWC 349A.

The radiative transfer simulations show that passive irradiated disks, which have the majority of their $3.78~\mu$m flux near their inner rim, cannot match the observations given MWC 349A's inclination ($75^\circ$).
For low MWC 349A luminosity ($\sim3\times10^4$ L$\odot$), the extent of the emission is much too small to match the squared visibilities (Figure \ref{fig-radmcres}).
For a higher stellar luminosity ($\sim8\times10^5$ L$\odot$), the asymmetric disk rim at larger angular separation causes a phase signal that is too large.
A rounded inner disk wall would produce a lower phase signal \citep[e.g.][]{2006ApJ...647..444M}.
This would be consistent with previous interferometric observations of Herbig Ae/Be stars, which could not be fit by models with simple vertical disk rims \citep[e.g.][]{2006ApJ...647..444M,2016ApJ...826..120M,2017A&A...599A..85L}.
However, even a perfectly symmetric ring (see Figure \ref{fig-symmring}) does not match the data, since the large inclination shortens the appearance of the disk on the sky.
This results in squared visibilities that fall off too quickly with baseline length.
Rounded rim models with an inclination of $\sim 48^\circ$ can match the data; however this is unlikely given previous constraints on the disk inclination from radio recombination line observations \citep[e.g.][]{1994ApJ...428..324R}.

In both the high and low luminosity case, reproducing the observations requires additional emission, and thus heating, at large radii.
The maser emission far from the star supports this scenario, since masers are often caused by shocks which would heat nearby gas \citep[e.g.][]{2016A&A...592A..31L}.
The presence of an ionized outflow \citep[e.g.][]{2011A&A...530L..15M} is also consistent with heating at large radii.
This extended emission may support a young age for MWC 349, since previous observations of Herbig Ae/Be stars suggest the presence of extended envelopes \citep[e.g.][]{2017A&A...599A..85L}.

\begin{figure}
\epsscale{1.2}
\plotone{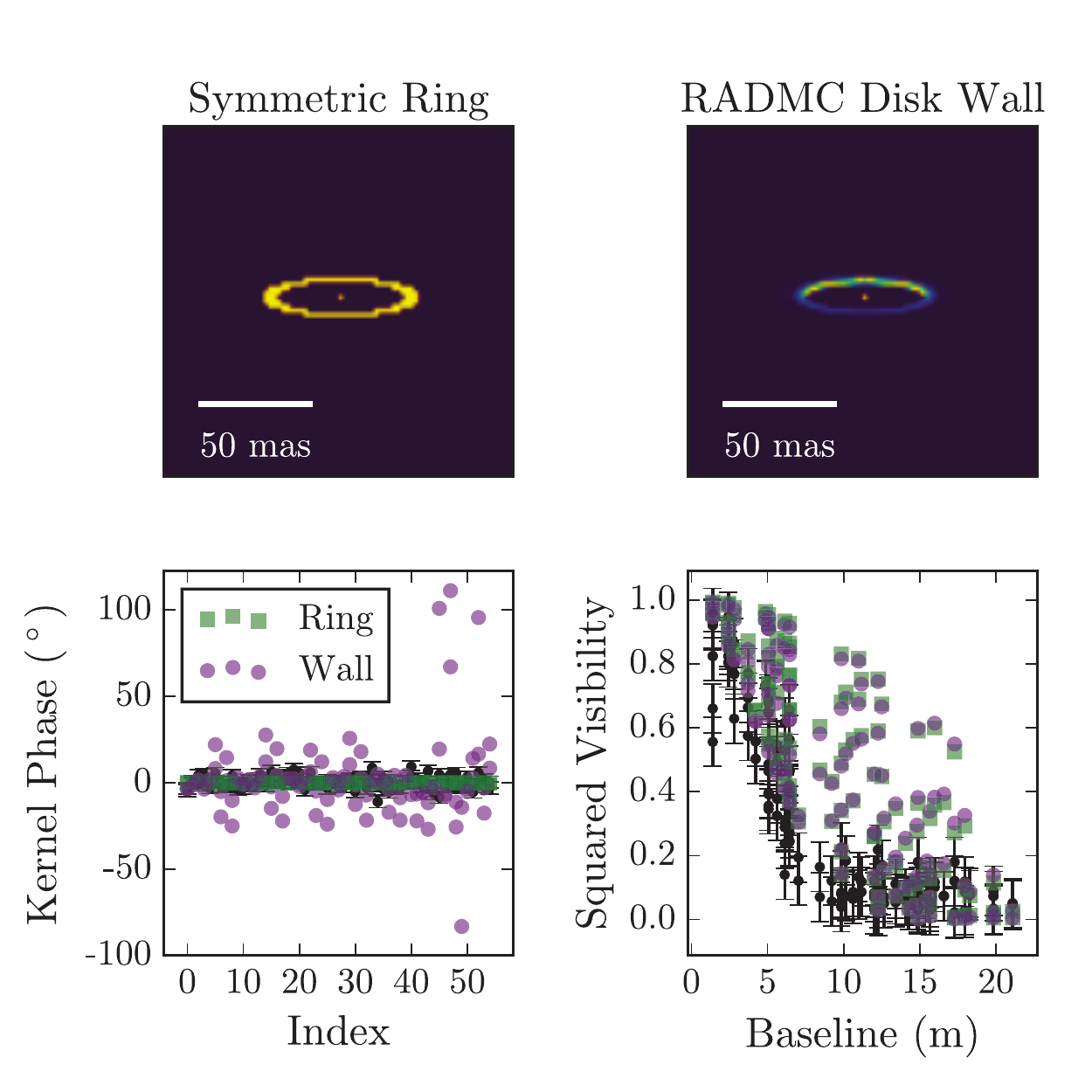}
\caption{\textbf{Top:} Left: Symmetric ring model illustrating the effect of a rounded inner disk rim. Right: \texttt{RADMC} radiative transfer model for a high luminosity MWC 349A. Both images have been rotated so the disk major axis is aligned with the x axis. \textbf{Bottom:} Kernel phases (left) and squared visibilities (right) for the model images, with the symmetric ring model in green and the higher stellar luminosity radiative transfer model in purple. Black points with error bars show the observations.}
\label{fig-symmring}
\vspace{10pt}
\end{figure}

\subsection{A Tight Binary?}
Some studies suggest that MWC 349 may be a hierarchical triple, where A is a close-separation binary surrounded by a circumbinary disk \citep[e.g.][]{2012A&A...541A...7G}.
Regular brightness variations with a period of nine years \citep{2000AJ....119.3060J} suggest that MWC 349A may indeed be a close binary system with an orbital separation of $\sim 13$ AU (7.7 - 10.8 mas depending on the distance estimate to MWC 349A). 
Given their resolution, previous infrared interferometric observations cannot rule out an embedded binary with a separation $< 28$ mas \citep{2001ApJ...562..440D}.
Our observations also cannot rule out a close-separation binary morphology for MWC 349A.
Model fits that include two point sources within the disk clearing can provide good fits to the data and do not tightly constrain the locations or fluxes of either inner component. 

\section{Conclusions}

We presented new, 23-meter baseline interferometric observations of MWC 349A from LBTI.
We fitted the data with geometric and radiative transfer models.
Geometric models with both skew and a compact component provided the best fit to the observations.
Models including an inner clearing constrain any disk hole to be less than $\sim14$ mas in radius.
The best-fit outer radii and skew parameters in the geometric models suggest the presence of a flat disk around MWC 349A. 
However, radiative transfer models of highly-inclined, passive irradiated disks cannot reproduce the observations and require additional heating at large radii.
The higher MWC 349A luminosity estimates require the presence of optically thick gas or refractory dust within the sublimation radius to match the compact infrared excess.
This scenario may support a young age for MWC 349.
In the low-luminosity case, determining the symmetry of the disk inner rim or detecting gaseous emission within the dust sublimation radius would help to constrain the age of MWC 349A. 
Making this distinction and placing constraints on possible close-in companions requires followup observations with increased sky rotation and higher resolution. 

\acknowledgments

This work was supported by NSF AAG grant $\#$1211329. This material is based upon work supported by the National Science Foundation Graduate Research Fellowship under Grant No. DGE-1143953. This material is based upon work supported by the National Science Foundation under Grant No. 1228509. Any opinion, findings, and conclusions or recommendations expressed in this material are those of the authors(s) and do not necessarily reflect the views of the National Science Foundation.

\bibliography{references}

\end{document}